%% file: main.tex
\documentclass[acmsmall]{acmart}

\AtBeginDocument{%
  \providecommand\BibTeX{{%
    \normalfont B\kern-0.5em{\scshape i\kern-0.25em b}\kern-0.8em\TeX}}}
\usepackage{csquotes}
\renewcommand{\mkbegdispquote}[2]{\itshape}
\setcopyright{rightsretained}

\acmJournal{PACMHCI}
\acmYear{2022} \acmVolume{6} \acmNumber{CSCW2}
\acmArticle{377} \acmMonth{11} \acmPrice{15.00}
\acmDOI{10.1145/3555102}
\received{July 2021}
\received[revised]{November 2021}
\received[accepted]{February 2022}

\copyrightyear{2022}
\acmYear{2022}
\setcopyright{rightsretained}

\acmJournal{PACMHCI}
\acmYear{2022} \acmVolume{6} \acmNumber{CSCW2} \acmArticle{377} \acmMonth{11} \acmPrice{} \acmDOI{10.1145/3555102}

\usepackage{etoolbox}
\makeatletter
\patchcmd{\maketitle}{\@copyrightpermission}{
   \begin{minipage}{0.2\columnwidth}
     \href{https://creativecommons.org/licenses/by/4.0/}{\includegraphics[width=0.90\textwidth]{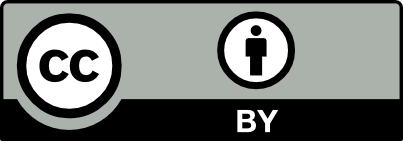}}
   \end{minipage}\hfill
   \begin{minipage}{0.8\columnwidth}
     \href{https://creativecommons.org/licenses/by/4.0/}{This work is licensed under a Creative Commons Attribution International 4.0 License.}
   \end{minipage}

   \vspace{5pt}
}{}{}

\makeatother

\usepackage{xcolor}
\usepackage{todonotes}
\begin{document}

\title{The Distressing Ads That Persist: Uncovering The Harms of Targeted Weight-Loss Ads Among Users with Histories of Disordered Eating}

\author{Liza Gak}
\email{lizagak@berkeley.edu}
\affiliation{%
 \institution{University of California, Berkeley}
 \streetaddress{South Hall}
 \city{Berkeley}
  \state{California}
  \country{USA}
  \postcode{94704}
  }

 \author{Seyi Olojo}
 \email{oolojo@berkeley.edu}
 \affiliation{%
   \institution{University of California, Berkeley}
  \streetaddress{South Hall}
   \city{Berkeley}
   \state{California}
   \country{USA}
   \postcode{94704}
   }
 \author{Niloufar Salehi}
 \email{nsalehi@berkeley.edu}
 \affiliation{%
   \institution{University of California, Berkeley}
  \streetaddress{South Hall}
   \city{Berkeley}
   \state{California}
   \country{USA}
   \postcode{94704}
   }

\renewcommand{\shortauthors}{Liza Gak, Seyi Olojo, \& Niloufar Salehi} 
\renewcommand{\shorttitle}{The Harms of Targeted Weight-Loss Ads Among Users with Histories of Disordered Eating}

\begin{abstract}
Targeted advertising can harm vulnerable groups when it targets individuals' personal and psychological vulnerabilities. We focus on how targeted weight-loss advertisements harm people with histories of disordered eating. We identify three features of targeted advertising that cause harm: the persistence of personal data that can expose vulnerabilities, over-simplifying algorithmic relevancy models, and design patterns encouraging engagement that can facilitate unhealthy behavior. Through a series of semi-structured interviews with individuals with histories of unhealthy body stigma, dieting, and disordered eating, we found that targeted weight-loss ads reinforced low self-esteem and deepened pre-existing anxieties around food and exercise. At the same time, we observed that targeted individuals demonstrated agency and resistance against distressing ads. Drawing on scholarship in postcolonial environmental studies, we use the concept of slow violence to articulate how online targeted advertising inflicts harms that may not be immediately identifiable. CAUTION: This paper includes media that could be triggering, particularly to people with an eating disorder. Please use caution when reading, printing, or disseminating this paper. 
\end{abstract}



\maketitle

\input{1-intro}

\input{2-relatedwork}

\input{3-method}

\input{4-results}

\input{5-discussion}
\input{6-conclusion}

\section*{Acknowledgements}
We would like to extend our deepest gratitude to our interview participants for sharing their insights and experiences. 
Additionally, we would like to thank both the Center for Technology, Society and Policy and the Algorithmic Fairness and Opacity Group at the University of California, Berkeley for their generous support and funding. We would also like to thank the participants and organizers at the What Can CHI Do About Dark Patterns? workshop at CHI 2021 for helpful feedback on earlier versions of this work. 

\bibliographystyle{ACM-Reference-Format}
\bibliography{main}

\end{document}

%% file: 1-intro.tex
\section{Introduction}

Targeted ads have raised privacy concerns \cite{ullah2020privacy}, as well as ethical concerns over surveillance and the use of persuasive technologies  \cite{saari2004psychologically, harris2012protecting}. In this paper, we investigate the harms of online advertisements that target users' vulnerabilities. Specifically, we focus on the harms of weight-loss ads targeted to people with histories of unhealthy body stigma, dieting, and disordered eating. By interfacing with vulnerable communities, we aim to understand the nature of the harms and how users develop strategies to mitigate them. Specifically, we ask: 
\begin{itemize}
\item How are people with histories of unhealthy body stigma, dieting, and disordered eating harmed by targeted weight-loss ads? 
\item How do they mitigate the reach and harm of targeted weight-loss ads? 
\end{itemize}
We conducted a series of interviews with 21 individuals with histories of unhealthy body stigma, dieting, and disordered eating. One challenge with studying daily, repeated online activity is that it is sometimes difficult for users to recognize when they are harmed. To overcome this challenge, we identified online communities that were already having collective discussion about the harms of diet culture. Therefore, we recruited participants from Intuitive Eating communities on Reddit, an online forum with interest-based groups. Past research has shown that online communities serve as safe spaces for social support and information sharing \cite{devito2019social, andalibi2017sensitive, de2014mental, mcandrew2020you}. We recruited from these online communities and asked participants about their experiences with targeted weight-loss ads.

 We identify three features of targeted advertising that cause harm: the persistence of personal data that can expose vulnerabilities; over-simplifying algorithmic relevancy models; and design patterns encouraging engagement that can facilitate unhealthy behavior. Participants felt that weight-loss ads strategically targeted individuals who have interacted with weight-loss content, creating unique risks for people with histories of disordered eating. Even when participants were pursuing content related to Intuitive Eating, they reported that the targeting algorithm was unable to differentiate between weight-loss content and Intuitive Eating content. Participants reported that the personalized nature of these ads manipulated user behavior, and resurfaced prior destructive habits. 
 
We found that participants engaged in a sense-making process to recognize weight-loss ads as harmful and, in doing so, changed their behaviors to mitigate the harms of the ads. By reflecting on their personal vulnerabilities, and the unscientific health claims made in weight-loss ads, we found that the participants recognized and engaged in sense-making around the harm of targeted weight-loss ads. Weight-loss ads complicated participants' relationship with food and exercise, and reinforced low self-esteem. Finally, we found that our participants creatively mitigated and resisted the harm of online targeted weight-loss ads. However, they also emphasized the limitations of these actions to meaningfully repair harm or protect themselves and others from future harm.

We draw parallels between the harms of online targeted weight-loss ads and the literary scholar Rob Nixon's concept of slow violence, which captures gradual, delayed harm over time \cite{nixon2011slow}. Slow violence builds on the concept of structural violence, but centers temporality as a vehicle for obscuring and prolonging harm \cite{galtung1969violence}. Since ads are seldom regarded as a site for violence, the harms experienced due to targeted weight-loss ads are often difficult to trace. The act of naming this violence as ''slow'', especially within the context of targeted ad delivery, highlights the role of gradual repetition as a significant form of violence. We argue that targeted weight-loss ads reify existing social norms of thinness. In this analysis, we introduce slow violence as an applicable framework for the CSCW community in order to both broaden and complicate understandings of harm via socio-technical systems. We hope that in future work researchers will design, implement, and evaluate collective solutions to the harms of targeted advertising, and collaborate with harmed groups to respond to their temporal and localized concerns.

\section{Background: Intuitive Eating \& Diet Culture}
In this section, we provide background on Intuitive Eating and disordered eating, since we recruited participants from online Intuitive Eating communities. Additionally, we provide context to the historic harms of diet culture and advertising. 

Weight-loss advertising has been rampant since before technologies of ad targeting and digital marketing existed. Sociologist Sabrina Strings argues that racism and colonialism are at the foundation of Western diet culture, and that fatphobia functions as a means of using the body to validate race, class, and gender prejudice \cite{strings2019fearing}. Furthermore, Jean Kilbourne, an expert on gender in advertising, argues that advertising is a manifestation of the way in which neoliberal capitalism poses consumption as an answer to structural trauma and insecurity. Kilbourne explains that within Western cultural norms around self-control and hedonism, body image is a significant vector for social capital. The pervasive message of dieting, described by Kilbourne as ``the modern self-purification ritual,'' stays constant throughout time, space, and media \cite{kilbourne_still_1994}. 

A significant obstacle to recovery is that eating disorders, namely anorexia nervosa, are ``egosyntonic'' meaning that people who suffer from eating disorders value their symptoms, which hinders their recovery \cite{gregertsen2017egosyntonic}. The egosyntonic nature of eating disorders is important to note, especially when considering how vulnerable individuals may engage with weight-loss ads as a manifestation of the symptoms of their eating disorders.

Developed by dieticians and nutrition therapists, Intuitive Eating is an eating philosophy associated with positive body image, where individuals learn to eat in response to physical hunger and satiety cues, as opposed to eating in response to emotional or situational signals. An important part of recovery from disordered eating patterns, Intuitive Eating emphasizes that individuals cultivate awareness of when they are hungry or full, which is suppressed in people who chronically diet or engage in disordered eating \cite{tribole1995intuitive, wood2010but, tylka2006development, koller2020body}.

%% file: 2-relatedwork.tex
\section{Related Work}
In this work, we analyze the effects of online targeted advertising on a specific marginalized population, individuals with histories of unhealthy body stigma, dieting, and disordered eating. Our research builds on two areas of related work: online behavioral advertising, and harms \& justice in socio-technical systems. 
\subsection{Online Behavioral Advertising}

Personalization and behavioral targeting are effective at achieving high user engagement and response rates \cite{oberoi2017technology, sahni2018personalization, maslowska2016all, montgomery2009prospects, shanahan2019getting}. The advertising technology industry uses personalization, a form of data extraction, to collect user behavioral data in order to increase monetization \cite{zuboff2015big}. A majority of ad categories are behaviorally targeted \cite{liu_adreveal_2013} which can increase click-through rates up to 670\% \cite{yan2009much}. Furthermore, health ads rely on re-marketing, meaning that ads are delivered frequently to users expressing specific interest in health-related products \cite{liu_adreveal_2013}. We build on this work by analyzing the effects of engagement with targeted weight-loss ads on people with histories of disordered eating.

    Algorithmically-driven ad delivery that prioritizes ``relevance'' disadvantages both users and advertisers \cite{ali_ad_2019}. One study concludes that even when the target audience and budget are held constant, Facebook’s ad-delivery algorithm has significant control over the audience that an ad is delivered to \cite{ali_discrimination_2019}. In the context of political ads, ``relevant'' ad delivery limits users' exposure to diverse viewpoints, while advertisers face prohibitively expensive costs to deliver ads to users that the platform believes are aligned with the views of the ad. Research on the harms of ad targeting has mostly focused on targeting due to users' political interests \cite{liberini2020politics, bennett2019data, bennett2020understanding, pariser2011filter}, but targeting based on users’ interests presents specific types of harms, especially when users' vulnerabilities are categorized as interests \cite{cotter2021reach, keller2018toward}. We aim to critically examine the emphasis on ``relevance'' with respect to ad delivery about sensitive health topics, arguing that targeting algorithms over-simplify users' interests even when parameters for targeting are broad.

	Mental models and folk theories influence online behavior and attitudes around targeted ads \cite{eslami2016first, devito2018people, devito2017algorithms}. In the context of online behavioral advertising, Yao et. al. identify distinct folk models for how people understand online behavioral advertising; they argue that participants’ understandings of online behavioral advertising varied by three elements: who tracks users’ information, where the information is stored, and how the ads are delivered \cite{yao2017folk}. Understandings of privacy, particularly with respect to online behavioral advertising, are situated in legal, historical, and cultural contexts \cite{wang2016examining}. Overall, users have strong negative reactions to personal information being displayed in targeted ads \cite{hanson2020taking, ur2012smart, zeng2021makes}. Because of the lack of transparency in algorithmically-driven behavioral advertising, advertisers have begun communicating algorithmic processes to users in order to alleviate information asymmetry \cite{eslami_communicating_2018}. This research opens the opportunity to investigate how users perceive advertising based on sensitive health data.

    Ad targeting algorithms deliver stereotypically gendered content based on users' perceived gender, even when advertisers fail to specify gender in their targeting criteria \cite{ali_discrimination_2019}. In a series of experiments, Ali et. al. conclude that even when the target audience and budget are held constant, Facebook’s ad delivery algorithm has significant control over the audience that an ad is delivered to. In this study, they observe skewed delivery across racial and gender lines, leading to potentially discriminatory effects \cite{ali_discrimination_2019}. Broadly speaking, digital technologies related to exercise and dieting, such as fitness trackers, construct gendered norms \cite{gupta2020gender}, while marginalizing the most vulnerable users \cite{cifor2020gendered, simpson2017calorie, spiel2018fitter}. This frames our inquiry into online targeted weight-loss ads, since dieting and weight loss are constructed as highly gendered practices \cite{mcphail2009tubby, bordo2004unbearable, temple2015fat}. 

\subsection{Harms \& Justice in Socio-technical Systems}

We contribute to a body of CSCW scholarship applying understandings of social justice to algorithmic harm. CSCW scholarship has opened the area for discourse around the relationship between social justice and technology \cite{fox2017social, star1999layers, chancellor2019relationships, fox2017imagining, kumar2018uber}. Social justice addresses how people experience oppression, and considers how situated context and social issues affect experiences of harm \cite{fox2016exploring}. The categorization and classification that is inherent to online targeted advertising abstracts identities, thereby creating new forms of violence \cite{dencik2018conceptual}. We build on previous work by analyzing how targeted advertising categorizes and harms marginalized groups.

Pater and Mynatt define digital self-harm as ``the online communication and activity that leads to, supports, or exacerbates nonsuicidal, yet intentional harm or impairment of an individual’s physical wellbeing'' \cite{pater2017defining}. Eating disorder communities are a prime example of a community that engages in indirect self-harm; for instance, online pro-eating disorder communities exchange thinspiration, which are photos, videos, and prose intended to inspire eating disorder symptoms. We conceive of vulnerable people engaging with targeted weight-loss ads as an example of digital self-harm. By interrogating the persuasive design of weight-loss ads, we extend scholarship on dark patterns, user interface design choices that coerce users into making potentially harmful decisions to the benefit of an online service \cite{mathurdarkpatterns, gray2018dark, gray2021end}. While we don’t see the relationship between disordered eating and targeted weight-loss ads as causal, we argue that the dark patterns of targeted weight-loss ads cause harm by capitalizing off of users’ already existing vulnerabilities. While weight-loss ads feature design patterns \cite{alexander1977pattern} that increase engagement, there is great potential for weight-loss ads to be framed as coercive dark patterns that cause significant harm, and in some cases even digital self-harm. 

Previous research has examined the relationship between digital technology and the disordered eating community \cite{chancellor2016thyghgapp, feuston2020conformity, chancellor2016quantifying, pater2021charting}. Social media can contain communities that encourage disordered eating behaviors, rather than identifying such behaviors as harmful \cite{chancellor2016quantifying, pater2021charting}. Researchers have found that online content promoting the``thin ideal,'' such as weight-loss ads, promotes behaviors associated with eating disorders \cite{cavazos2020examining}. One study demonstrates that an increased use of social media leads to an increased incidence of anorexia and internalization of the thin ideal \cite{turner2017instagram}. 

As we acknowledge that the design of embodied technologies implicitly enforces bodily norms and assumptions \cite{spiel2021bodies}, we interrogate the harms of unhealthy body ideals central to targeted weight-loss advertising. Building on prior work, we highlight the unique needs of vulnerable people with histories of disordered eating.

%% file: 3-method.tex
\section{Method}

Our goal in this research is to understand how and why targeted ads harm people; in particular, we study how targeted weight-loss ads harm people with histories of unhealthy body stigma, dieting, and disordered eating. We conducted an interview study to observe peoples' sensemaking processes around targeted weight loss ads, and to capture how they anticipate, perceive, and respond to these ads. The challenge with studying harms that may seem small individually but accumulate to have harmful effects is that the people who are harmed may not immediately notice or identify the harm. For this reason, we decided to prioritize sampling from communities that have collectively discussed these issues and had already engaged in sense-making about the harms of targeted weight-loss ads prior to our study. We focus on the Intuitive Eating community on Reddit, which consists of people repairing their relationships with food and exercise. From this community, we recruited individuals with histories of unhealthy body stigma, dieting, and disordered eating. Through our interviews, we study the needs of harmed users, their perceptions of harm, and what strategies participants take ``on the ground'' to mitigate harm. 

\subsection{Study Design} 
We conducted 21 semi-structured interviews with adults (18+) with histories of unhealthy body stigma, dieting, and disordered eating, primarily from Intuitive Eating communities on Reddit.

\subsubsection{Intuitive Eating Communities}
We recruited participants from Intuitive Eating communities on Reddit, composed of the subreddits r/IntuitiveEating (21.6k+ members), r/EatingIntuitively (4.1k+ members), and r/AntiDiet (7.3k+ members), where members were already discussing their experiences with the harms of weight-loss culture and advertising. Members of the Intuitive Eating community subscribe to the Intuitive Eating framework, which are eating philosophies valuing every individual’s choice to listen to their own hunger cues and reject diet mentality \cite{tribole1995intuitive}. The community opposes diet culture, believing that diets are a starting point for disordered eating. Members of this community engage in information and resource exchange, and provide social support to one another. Intuitive Eating provides a path to recovery from disordered eating, as recommended by medical professionals \cite{tribole1995intuitive, wood2010but, tylka2006development, koller2020body}. We considered this community since its members have drawn attention to the harms of targeted weight-loss ads, and have collectively shared their reactions to ads as well as strategies for mitigation. 

\subsubsection{Representation in Recruitment}

The prevalence of disordered eating differs across ethnic and racial groups \cite{powell1995racial, walcott2003adolescents}; therefore, we wanted our sample to reflect the diversity of those experiences. To expand the diversity of our participant pool, we conducted a second round of outreach aimed at recruiting participants of color, because the majority of participants in the first round were White. We posted again to the online communities emphasizing that we were recruiting eligible BIPOC (Black, Indigenous, and People of Color) participants, where we were able to recruit a more diverse participant pool. However, we were not able to reach eligible and interested Black participants in both the first and the second round of outreach. Our second author posted to a group she was already a part of, r/BlackLadies, an online community on Reddit for Black women. In this additional round of recruitment, we clarified eligibility criteria to recruit adults familiar with the Intuitive Eating philosophy. We made sure that those recruited outside of Intuitive Eating communities were already familiar with the philosophy to keep consistency across our participant pool. In the future, we aim to reach a more diverse audience of participants that better reflects the groups of people vulnerable to targeted weight-loss advertising.

\begin{figure}[htp]
    \centering
    \includegraphics[width=0.75\textwidth]{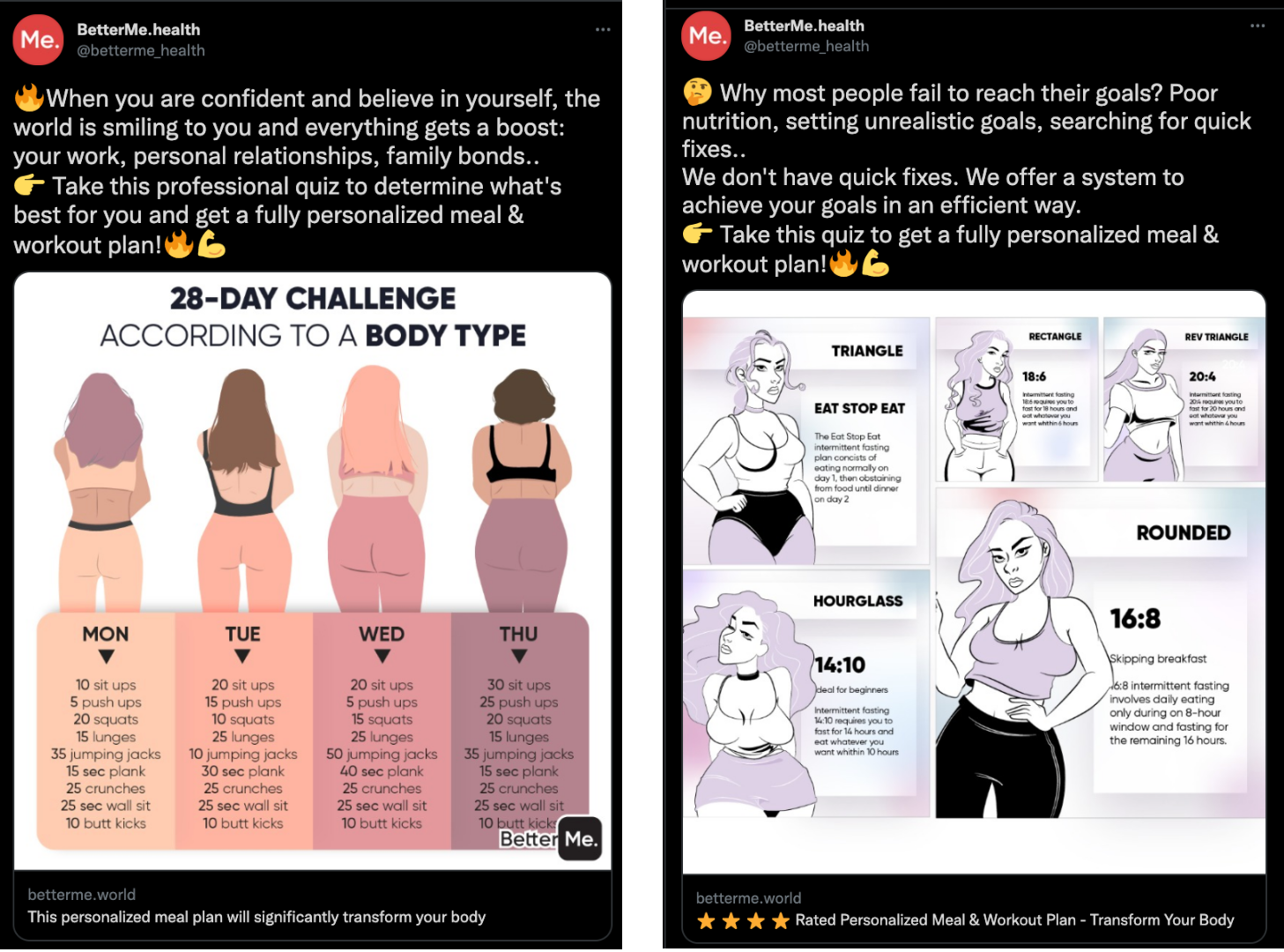}
    \caption{Typical weight-loss ads that we used in interviews. }
    \label{fig:ads}
\end{figure}

\subsubsection{Interview protocol}
During our interviews we showed participants a typical online weight-loss advertisement with the goal of understanding their reactions and mental models after seeing the advertisement, as seen in Fig. 1. We wanted to put the participants in a space for critical reflection, which is difficult to accomplish without an example of a weight-loss ad. We chose these particular ads as examples because while they exemplified a typical personalized weight-loss ad with their language, imagery, and potential for engagement, they were not gratuitous. Analyzing an example of a weight-loss ad also solidified a common understanding of weight-loss ads, and gave participants a more concrete opportunity to describe how they anticipate, perceive, and react to these ads. We asked why they thought they might have received the ad. While we did not anticipate adverse reactions because participants see such ads on a regular basis, we emphasized that this part of the interview was optional. We used this example as a jumping off point for discussing participants’ experiences with targeted ads.

In addition to understanding the nature of the harm, we were interested in learning what mitigation strategies people currently use or could potentially use. We presented six storyboard comics, where each comic sketches out a common strategy for resisting distressing ads, such as reporting the ad, commenting on the ad, or sharing the ad to their Intuitive Eating community. Fig. 2 illustrates one of the storyboard comics used during the interviews. Then, we asked participants to identify which strategies they had taken, and what felt most effective. The comics acted as a probe \cite{gaver1999design} for brainstorming other routes for resistance, since participants were able to compare and contrast the strategies proposed in the comics. Furthermore, the storyboards helped expand the axes of brainstorming by introducing both individual strategies and possibilities for collective action. We asked participants to reflect on the personal value of the Intuitive Eating community, in order to get a sense of potential collective action opportunities.

We conducted interviews from December 2020-March 2021. We paid participants a \$30 Visa or Target gift card for 45 minutes of their time. We recorded the interview as per the participants’ approval, and later anonymized and transcribed it. To ensure the privacy and anonymity of participants, we identify participants in this paper by unique identifiers P1 through P21. The Intuitive Eating community is a global community, with members from multiple continents. Because we wanted to reflect geographic diversity of the Intuitive Eating community, we interviewed people affected by weight-loss ads from around the world. While a majority of the participants reside in the United States, some participants were from Norway, Australia, New Zealand, and Canada, which is reflected in the demographic table. Our participants were 25.8 years old on average, with an age range from 19 to 35. The majority of our sample identified as female, with 17 female participants, 3 non-binary participants, and 1 who preferred not to respond. This gender distribution is to be expected because women are more frequently targeted and harmed with weight-loss advertisements \cite{ali_discrimination_2019}.

\begin{center}
\begin{tabular}{||c c c c c||} 
 \hline
  Identifier & Age & Race and Ethnicity & Gender & Country \\ [0.5ex] 
 \hline\hline
 P1 & 33 & White & Female & Canada \\ 
 \hline
 P2 & 19 & White & Female & USA \\
 \hline
 P3 & 35 & White & Female & USA \\
 \hline
 P4 & 23 & Middle Eastern & Non-binary & USA \\
 \hline
 P5 & 21 & White & Female & USA \\
 \hline
 P6 & 19 & Asian & Female & USA \\
 \hline
 P7 & 23 & Pacific Islander + Filipino & Prefer not to respond & USA \\
 \hline
 P8 & 24 & White & Female & USA \\
 \hline
 P9 & 34 & White & Female & USA \\
 \hline
 P10 & 25 & Biracial (South Asian + Latinx) & Female & USA \\
  \hline
 P11 & 25 & White & Female & USA \\
 \hline
  P12 & 21 & Black & Female & USA \\ \hline
 P13 & 23 & Biracial/Mixed & Female & Norway \\
  \hline
   P14 & 35 & Black & Female & USA \\
  \hline
   P15 & 19 & Asian & Non-binary & USA \\
  \hline
   P16 & 21 & Indian (Asian) & Non-binary & USA \\
  \hline
 P17 & 30 & White & Female & Australia \\
 \hline
 P18 & 24 & New Zealand European/Pākehā & Female & New Zealand \\
 \hline
 P19 & 33 & White & Female & USA \\
 \hline
 P20 & 34 & White & Female & USA \\
 \hline
 P21 & 20 & Asian-American & Female & USA \\
 \hline

\end{tabular}
\captionof{table}{Participant Demographic Information} 

\end{center}

\begin{figure}[htp]
    \centering
    \includegraphics[width=0.75\textwidth]{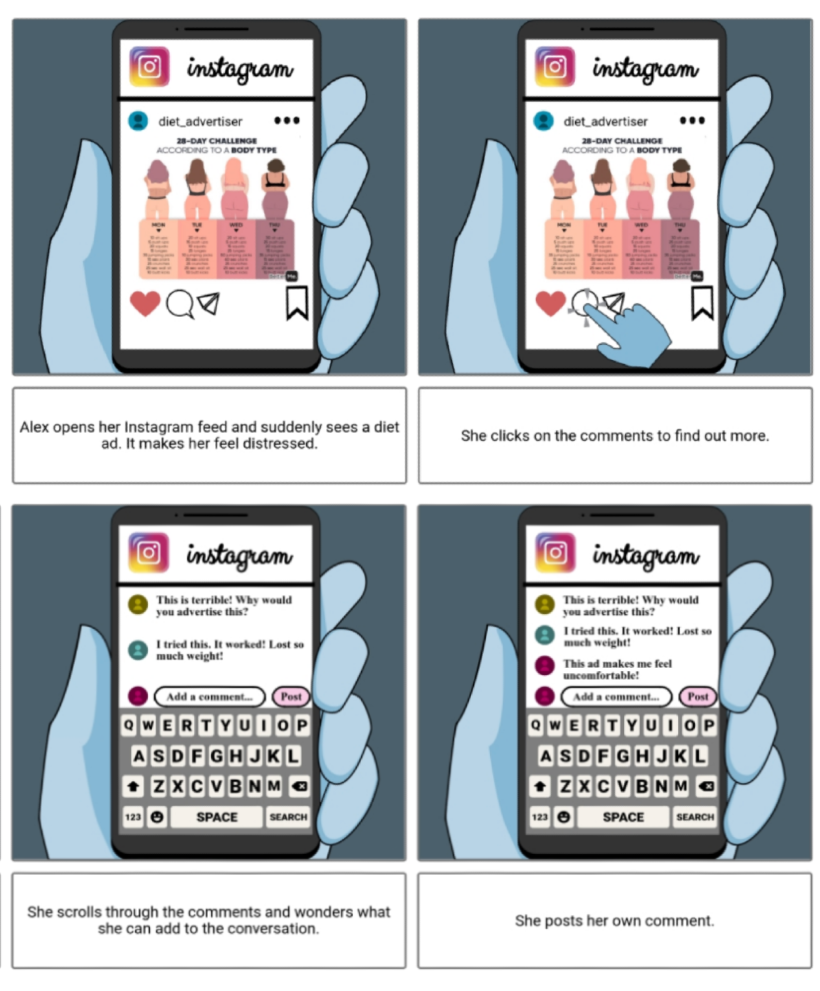}
    \caption{Example of a storyboard that we used as a probe \cite{gaver1999design} during interviews. In this scenario the user engages in a conversation about the harms of the ad in the comments section. Other scenarios include ignoring, reporting, muting, or clicking on the ad, and sharing the ad with the Intuitive Eating community. }
    \label{fig:galaxy}
\end{figure}

\subsection{Ethical considerations}
Because we asked participants about sensitive topics, we were mindful to ensure the comfort and safety of our participants and we took multiple steps to ensure ethical compliance. We received IRB approval from our university which classified this study as minimal risk because we did not expose participants to harms beyond what is ordinarily encountered in daily life. We considered how participants might be harmed during the interview and took steps in our study design to minimize potential discomfort. During the recruitment process, we made participants aware that we would discuss examples of weight-loss advertisements during the interview, and would be discussing their experiences with targeted advertising more broadly. Before each interview formally began, we explained to the participants that their participation in the research study was entirely voluntary,  that they could decline answering any questions, and that they could stop the interview at any time and still be compensated for their participation. If an adverse reaction would have occurred, we would have immediately stopped the interview and referred the participant to relevant resources. 

Before presenting a weight-loss advertisement to participants, we asked if they consented to seeing a weight loss advertisement. After they gave verbal, affirmative consent to seeing a weight loss ad, we gave them the opportunity to choose between two typical advertisements and discuss their reflections of one of them. The weight-loss advertisements we chose lacked gratuitous imagery, pictures of humans, and overtly dangerous language. Additionally, by providing options about what to discuss, we gave participants the opportunity to avoid any potentially triggering topics.

\subsection{Analyzing data}
To analyze the results the first and second author adopted an inductive-interpretivist approach. With this approach we made sure to draw a majority of the analysis from the data itself, rather than building off of previously existing theories. We used MAXQDA, a qualitative data analysis software to code recurrent key terms, themes, and concepts from all 21 transcripts. In the initial review of the data, we analyzed the transcripts using line by line coding. Then, we discussed and refined emerging keywords related to RQ1 and RQ2. Examples of such keywords include ``unhealthy body ideals'' and a ``concern for younger population.'' In a second review of the data, we individually analyzed the transcripts using keywords from the first review of the data as the set of codes. After this second review, we agreed upon a broader set of themes based on the frequencies of keywords within the transcripts. These themes included ``lack of scientific rigor,'' ``trauma,'' and ``personalization.'' In order to draw relationships between the keywords and identify broader emergent themes, we created a concept map using Mural, a visual collaboration platform. Salient keywords were grouped under larger themes, such as ``persistence of personal data,'' ``logics of relevancy,'' ``user agency as resistance.'' In the refinement of these themes, we went back to the coded transcripts as needed.

%% file: 4-results.tex
\section{Results}
In this paper, we identify three major ways in which targeted advertising harms vulnerable people: 1) the persistence of sensitive personal data, 2) over-simplifying algorithmic relevancy models, and 3) design patterns that nudge toward unhealthy behavior. Next, we explain how our participants engaged in sense-making and described the effects of the ads on themselves. Participants identified the harm of targeted weight-loss ads by reflecting on their own past vulnerabilities, the pseudo-scientific content in the ads, and the emphasis of the ads on individual responsibility. We describe how targeted weight-loss ads harmed our participants, including emotional and physical harms. Finally, we describe tactics that our participants used to mitigate the harm including blocking and reporting the ads as well as collective deliberation around the harms of the ads. Our participants used varied approaches in resisting the harmful effects of weight-loss ads, but were aware of the limitations.

\subsection{Sensitive Data, Relevancy Models, and Design Patterns Cause Harm in Targeted Ads}
Our participants theorized that the persistence of their sensitive, personal data exposes their vulnerabilities to targeting algorithms. Second, algorithmic relevancy models over-simplify interest in a topic. Finally, design patterns that increase engagement increased the potential for harm. 
In the following sections, we detail each of these mechanisms. 

\subsubsection{Persistence of sensitive, personal data exposes vulnerabilities}
Our study participants believed that they were targeted with weight-loss ads because of their search histories as well as their demographic characteristics, such as age and gender. Online targeted ads rely on a large amount of personal data to decide who might be interested in which ad. However, peoples' personal data holds sensitive information that can lead to exposing and inadvertently exploiting their vulnerabilities.

The online activity of people with histories of disordered eating is imbued with weight-loss content, including their search queries, browsing history, and other online behavior. Personal data is highly connected and shared across platforms \cite{zuboff2015big}.
Participants theorized that targeting algorithms are aware of their personal and psychological vulnerabilities. In particular, they speculated that weight-loss ads leverage data that points to their personal histories of disordered eating.
\begin{displayquote}
They probably know my history of eating disorder and self-loathing, and that kind of stuff. They've probably seen Google searches of calorie [sic], substitutions, or whatever, that kind of stuff. (P3)
\end{displayquote}
Even when participants were practicing Intuitive Eating, they perceived that they were being targeted based on past data that was a symptom of their disordered eating. P19 explains that before they pursued Intuitive Eating, they searched content about dieting and weight loss. P19 speculates that they are a target for weight-loss ads because of their past queries: 
\begin{displayquote}
Well, I probably am getting targeted to get ads like this because I've Googled a lot, like how to lose weight, dieting, keto diet. What diet should I do? I mean, more so before I kind of really got serious about Intuitive Eating. I was very obsessed about diets and Googling that. (P19)
\end{displayquote}
Participants expressed that the personal data that weight-loss ads take advantage of is not just from recent searches, but rather persists over time. For instance, one participant started pursuing weight loss in elementary school, and eventually developed an eating disorder in high school. She felt that weight-loss ads were successful in engaging her because they targeted her vulnerabilities via data she produced a decade ago. 
\begin{displayquote}
Who I am now at 25 versus who I was at 15 are two very different people, especially with the things that I was struggling with at 15 because that was when the eating disorder behavior really became a problem, and still being fed to me even now that I have bad days, but it's not an all encompassing thing anymore. So, the fact that it just keeps coming up is a little tiring and a little bit [\ldots]  I almost want to say insulting that it's still working. These are still the things they think are going to get me. (P11)
\end{displayquote}

Participants believed that because they had searched for content related to disordered eating, whether while struggling with disordered eating or while pursuing Intuitive Eating, they were targets for weight-loss ads. Therefore, interests such as diets that once felt relevant can be brought up as triggering reminders of past issues. Targeted advertising relies on a wealth of user data to target ads, but while doing so reveals intimate vulnerabilities. 

\subsubsection{Relevancy models over-simplify users' interests}

We found that targeting algorithms construe broad audiences as ``relevant'' user groups for weight-loss content. Our participants believed that they were targeted because of persistent, sensitive personal data, their demographics, and their social networks -- all of which they felt they could not escape. Some participants reported that the targeting algorithm was unable to differentiate between weight-loss content and Intuitive Eating content. The participants we recruited from online Intuitive Eating communities were largely still interested in food and exercise content, but framed in ways that are appropriate for those with histories of disordered eating. Participants expressed frustration that interest in Intuitive Eating content, which opposes diet culture, is interpreted by the algorithm as an appropriate audience for weight-loss ads. Many participants responded with frustration as their genuine interests in Intuitive Eating felt erased.
\begin{displayquote}
I mean, I know with a lot of Intuitive Eating accounts or blogs, people think that Intuitive Eating is just another kind of diet and it's not, obviously, but there's so much diet content and blogs and advertisements and stuff like that are related to dieting and ``health.'' So because I follow a few accounts or just people in general that are Intuitive Eating adjacent, I could see ad targeting specifically thinking like, oh, you're in[to] health and fitness and dieting, so this [weight-loss] ad would align. (P16)
\end{displayquote}

The targeting algorithm conflates concepts that seem relevant on the surface, and even use the same language, but are actually different, and even oppositional ideas. In doing so, the algorithm pushes toward dominant narratives on a given topic and makes it harder for people to oppose or seek alternatives. Even though Intuitive Eating and anti-diet content stands as an overt critique of diet culture, participants theorize that the presence of the word ``diet'' and other related keywords make weight-loss ads relevant in the eyes of the algorithm. 
\begin{displayquote}
I still get targeted diet stuff just by being in Intuitive Eating communities because I think there's something in my data that says that I'm interested in nutrition and oftentimes I think when people think nutrition, they think weight loss or something with weight and healthy and these are the things that come up when you hear those words. I was scrolling on TikTok and because I liked a couple of videos of body neutral fitness workers on TikTok, I'm getting a lot of diet stuff now. All of a sudden a lot of people are like, ``Here's how you can lose that belly.'' I'm like, ``Oh, that was a very fast turnaround,'' even though there's this huge difference between nutrition and weight loss. (P11)
\end{displayquote}

Most participants felt that their demographic data, namely age and gender, played a significant role in receiving weight-loss ads. As P5 explains, once users indicate interest in topics ``relevant'' to dieting and weight loss, and are part of ``relevant'' demographics, they are inundated with weight-loss ads.  
\begin{displayquote}
They look at your demographics of [sic] you're a young woman and you're looking into stuff for health or I'll look up [\ldots] Oh, god. If you look up any stretching or yoga, immediately you'll get ads of yoga for weight loss, targeted yoga to shrink your stomach. I'm like, ``No, I just want to stretch.'' (P5)
\end{displayquote}
It’s also important to note that how targeting algorithms read demographic data does not always align with how participants saw themselves -- for example, P16 is non-binary but is still deeply affected by the misogyny of diet culture, ``I'm non-binary but obviously, I was socialized as a girl and most of my friends are women or feminine people. So it's just kind of a thing that affects everyone, basically, in a negative way.'' Not only is browsing data persistent in its targeting, but demographic data can be also inflexible.  

Participants perceived that targeting algorithms delivered ads based on the content of who they were connected to online. Even if users weren’t actively seeking out content related to weight loss, they still reported seeing ads based on who they were following. When asked why she thinks she receives weight-loss ads on Twitter, P21 theorized that the relevancy models targeted ad content based on an individual’s social connections. Participants cited the role of following influencers and content creators as highly influential in determining ad content. 
\begin{displayquote}

I think the Twitter algorithm, especially, does a good job of looking at people you're following, and if you're following people who are trainers or also post a lot about losing weight, that might be a reason why I would get this ad. (P21)
\end{displayquote}

Targeting algorithms use models that over-simplify relevancy based on perceived interest, demographics, and social connections. This makes it hard for people who can be harmed by the ads to avoid them.

\subsubsection{Design patterns meant to increase engagement can encourage unhealthy behavior}

Targeted ads use design patterns to increase engagement, such as personalized language and interactive quizzes. These techniques may encourage destructive behavior (e.g. excessive exercise, restricting food consumption, or dieting) for people with histories of disordered eating. For instance, ads that use engaging design patterns offer seemingly individualized approaches that encourage unhealthy behaviors. 

\begin{displayquote}

On TV it seems like when you see an [weight-loss] ad like that, or at least for me, it's like, "Okay, this is just a silly ad." [On the other hand], when it's in your personal feed, it seems a little more tailored to you. It seems like I would probably pay attention to it more because it's right there; instead of just on the TV, it's just going to flip off in a second. Whereas when you're scrolling with the target[ed] ad, you can click on it and you can really get involved in it. (P3)
\end{displayquote} 

Popular weight-loss ads feature Buzzfeed-style quizzes where users input personal, biometric data to get seemingly tailored diets and exercise. Even though the results are based on arbitrary factors like sleeping schedule and body shape, participants emphasized that weight-loss ads were appealing since they claim to offer customized approaches to rapid weight loss. Some of our participants identified their engagement with weight-loss ads as \textit{symptoms} of their disordered eating. In this sense, engagement with targeted weight-loss ads becomes a form of ``digital self-harm,'' defined by Pater and Mynatt as online activities that cause impairment of a person's wellbeing \cite{pater2017defining}. Due to the egosyntonic nature of disordered eating, individuals value the symptoms of their disorder even though it is harming them, which makes them especially vulnerable to design patterns that nudge users toward higher engagement. 
Because quiz results were personalized, P13 described that preexisting vulnerabilities, like disordered eating habits, made opportunities for ad engagement more appealing. After showing an engaging weight-loss ad during the interview, P13 explained why they would be persuaded to interact with this weight-loss ad:

\begin{displayquote}
Of course, if you want to lose weight, you obviously would want to take the tests and figure out what this ad means is the best way for you to lose weight. (P13)
\end{displayquote}

Similarly, another participant described their experience with intermittent fasting as a result of a personalized ad with engaging design patterns. P12 explained that the quiz used shame to lure individuals into trying their approach and that the questions in the quiz were capitalizing off of their personal trauma.
\begin{displayquote}
I think an ad like this is actually how I ended up doing intermittent fasting [\ldots] I took a mini quiz. It showed me the results. And then I started doing more research on it. The questions in the quiz bring a lot of harm and trauma, like ``When was the last time you were at your ideal weight?'' [They ask] to make you feel like crap, [making me respond] that ``Oh I've never been at my ideal weight.'' It just as with this extra layer of shame again like ``What? One to two years ago? It's been that long since you've been at your ideal weight''? (P12)
\end{displayquote}

Overall, we found that targeted ads use design patterns such as personalized language and quizzes and challenges to increase engagement, but this also encouraged dangerous behavior related to disordered eating for our participants. 

In this section, we presented three central features of targeted ads that can cause harm.  We argue that persistent, sensitive personal data reveals users' vulnerabilities, relevancy models over-simplify users' interests, and design patterns can engage users to the point of digital self-harm.  In the next section, we describe the physical and emotional effects of this harm on our participants.

\subsection{Sense-making and the Effects of Targeted Weight-Loss Ads}
 Harmful targeted ads can be difficult to recognize because of the seemingly small impact of the harm. Unlike egregious forms of online harm such as harassment and doxxing, the harms associated with targeted ads may not be immediately obvious, even to the people who have been harmed. Therefore, we chose to study people in online Intuitive Eating communities, since members had already done some of the work of discussing their experiences and identifying and naming shared issues that they faced \cite{fraser1990rethinking}. Our participants identified that targeted weight-loss ads were harmful by reflecting on their own past vulnerability to them, the lack of scientific backing for the claims in the ads, and the emphasis of the ads on individual responsibility. 
 
 In this section we begin by describing our participants' sense-making process for identifying the harm and its sources. In the section that follows, we will describe our participants' attitudes to potential mitigation strategies.

\subsubsection{Reflecting on past vulnerabilities.}
Our participants, who share histories of unhealthy body stigma, dieting, and disordered eating, identified their past selves as susceptible to engaging and subsequently harming themselves with weight-loss ads.

\begin{displayquote}
I know that's not me anymore, but a past me, I feel like, would have definitely responded even more so to this. So, if a past self who's really wrapped up in the way I'm eating and the insecurity and the lack of self-confidence, and looking at this saying, ``Okay, this is what I should look like. This is what I need to do to look like this or else I'm not good enough. I'm not worthy. My self esteem is super low. I'm not going to be desirable to anyone.'' (P3)
\end{displayquote}

Therefore, our participants identified people who would be most harmed by targeted weight-loss ads by reflecting on their own past and by generalizing to others with similar characteristics such as loved ones or other young women in their lives. 
\begin{displayquote}
I think of the danger when I look back at my younger self or even when I think about my niece, who's eight years old [\ldots] The idea of her getting served ads about how she needs to change her body is just[\ldots]  I don't understand why we aren't all morally outraged about that. It seems so horrible and so mean and so unnecessary all at the same time. (P9)
\end{displayquote}

Participants also challenged normative definitions of who could engage in disordered eating, and in doing so, expanded who could be considered vulnerable to  weight-loss ads. 
\begin{displayquote}
In terms of the targeted ads, I think that it's very frustrating and to think of all the people who are also in my demographic and they're also being targeted by this [\ldots] Also, it's expanding. Right? More and more men are having eating disorder[s] and trans people are one of the most likely groups to have eating disorders. I get ads sometimes where it's like a woman on one side and a man on one side and it's like, ``Diets for the different genders. What you can eat for each.'' (P5)
\end{displayquote}

Overall, our participants reflected on their past selves, particularly when they were more vulnerable, to identify the harm perpetuated by weight-loss ads. Through a process of hypothetical thinking and recalling past experiences, participants identified people who might be most vulnerable. 

\subsubsection{Pseudo-scientific claims in ads.}
Most of our participants described the lack of scientific backing for the ads as a major reason for identifying them as harmful. The content in online targeted ads does not undergo a rigorous verification process and is therefore, often unsubstantiated by scientific evidence. 
Participants felt that ad content should be based on legitimate and reliable information, especially with respect to sensitive health topics such as weight loss. 
\begin{displayquote}
I feel annoyed. Maybe I would want these things to be truthful but if they were truthful, they wouldn't be making money if they were truthful. I think the other thing is just the untruthfulness but also, the preying on people who feel insecure about their bodies is just very annoying to me. I feel like maybe annoying isn't the best word. It's really harmful and it just feeds into people feeling worse about themselves. (P7)
\end{displayquote}

Although the ads assume authority through ambiguous language, our participants believed that the information lacks scientific rigor, and is unsupported by reliable scientific institutions or other forms of community knowledge. 

\subsubsection{Emotional Responses \& Physical Harms}
During the interview and particularly after we presented examples of weight-loss ads, participants reflected on their initial emotional responses after seeing the ads. Most commonly, participants reported feeling frustration, annoyance, and disgust after seeing weight-loss ads.

Participants identified weight-loss ads as reinforcing low self-esteem and lack of control. Because the ads encourage existing insecurities, participants claimed that the ads bolstered existing feelings of self-doubt and helplessness. After seeing an ad that displayed unrealistic body results due to unhealthy weight-loss regimens, P9 describes that the ad intensifies her insecurities. 
\begin{displayquote}
I could see this setting off [\ldots] that voice in your head that says I'm not good enough [\ldots] The ``I'm not enough'' voice. Essentially, the message here is I shouldn't be confident in myself because there's no such thing as doing enough. (P9)
\end{displayquote}
Participants connected being delivered weight-loss ads to deepening feelings of anxiety around food. P3 describes that the ad reinforces depression and anxiety with respect to food and self-control, encouraging cycles of binge eating. She emphasizes that weight-loss ads’ focus on food and body image makes her more self-conscious. 
\begin{displayquote}
You feel depressed if you can't stick to this ad [because it] makes you feel bad about yourself, which I feel is like depression. It can also make you feel super anxious all the time about the food you're eating and the way you're looking or how other people are thinking of you and all of that can lead to behavior. At least for me, during the day I would keep calorie counting and track of everything and around others, add it together and all this stuff. And then at night, I would binge eat and then I felt depressed and out of control and like I don't have any self-control and there's something inherently wrong with me. (P3)
\end{displayquote}

Participants explained that being delivered weight-loss ads was emotionally and mentally draining.
Participants perceived ads as triggering reminders of harm they experienced offline, as a result of fatphobia and stigma around disordered eating. 
\begin{displayquote}

I think the biggest thing is this constant reminder that that mindset is out there and I know for me, it's been a big thing of, is this still what other people see? And then of course, you'll have that one person that says something to confirm that out loud and it's just this debilitating [\ldots] I only have so much energy to mentally spend on this and emotionally spend on this and ``I don't want to spend it all on you" kind of thing and that's kind of what it feels like is going on. It's just completely deciding where to put my emotional energy for the day. (P11)
\end{displayquote}

Participants believed that targeted weight-loss ads cast individual responsibility on the user, which undermines the recovery efforts of people who have histories of disordered eating. One participant pointed out that because weight-loss ads target vulnerable populations, individuals become disillusioned and blame themselves for failed products. 

\begin{displayquote}
It really bothers me that this stuff is just not true and it's preying on people who already feel insecure about themselves. If they're following the [product] and don't get results, then they feel it's on them that their body doesn't look like that. I think that's really harmful. People blame themselves for not looking this way when it's not true that you just do this one simple thing and then you'll get this specific kind of body. (P7)
\end{displayquote}

In addition to emotional and psychological harm, participants reported physical harm as a result of targeted weight-loss ads. 

P13 shared her story of buying rapid weight-loss pills from an online targeted ad, and having medical reactions from the purchased product. She recalls that at the time, she wasn’t old enough to buy the product due to age limits; however, an older friend bought the pills on her behalf. She describes that she was very susceptible to weight-loss ads when she was younger. After taking the pills, she started to experience physical side effects that were damaging. 

Additionally, weight-loss ads have the potential to influence users’ physical exercise habits. Many targeted weight-loss ads promote physical exercise products, with emphasis on frequency and intensity of exercise. The unhealthy focus on exercise and food in weight-loss ads made it more difficult for our participants to adhere to the principles of Intuitive Eating, and impeded progress to recovery.
\begin{displayquote}
But I'm also trying to be conscious about my movement and exercise, and seeing if I feel like exercising. Seeing stuff like that kind of makes me feel ashamed, like oh, I should be exercising more. Then that kind of makes it harder for me to know whether I feel [like] exercising. (P19)
\end{displayquote}
Our participants found targeted weight-loss ads highly damaging and distressing. Not only do weight-loss ads pose emotional and affectual harms, but ads and their associated products also have the power to inflict tangible physical harm on targeted individuals. Next, we discuss how our participants practiced resistance.

\subsection{Agency and Mitigation Strategies in Reaction to Harmful Targeted Ads}
Our participants practiced agency in the face of targeted ads through a process of reflection and resistance. Because users experienced emotional distress due to triggering ads, they changed their online behaviors to mitigate the reach and harm of these ads. Resistance took many forms, from individual approaches such as muting, blocking, or reporting, to more collective efforts such as discussing the harm in the comments section of the ad.

 Interviewees reported that ads impeded recovery progress from disordered eating, and that they had to self-soothe after seeing disturbing ads. On the other hand, they saw their resistance to engaging with the ads, including ignoring or blocking, as a sign of progress in their Intuitive Eating efforts. They also reflected on the limitations of their approaches in preventing future harm to themselves and others.

\subsubsection{Individual approaches: muting, blocking, and reporting}

Participants sought to mitigate harm as a response to receiving weight-loss ads, and also resisted being targeted by weight-loss ads in the first place. Our participants sought ways to avoid seeing weight-loss ads on their social media feeds. The most accessible and widely used strategies were individual -- muting, blocking, and reporting. These strategies yielded little success. 
A significant number of participants also reported decreasing their social media activity to avoid ads, and using third-party ad blockers. However, P1 highlighted limitations to ad blockers, such as the fact that their ad blocker fails to extend to browsing on a mobile device or a TV. 

On social media platforms, users often have the option to mute or hide an ad, or indicate that they are not interested in given ad content. Our participants were aware that an algorithm determines what ads to show them and they attempt to negotiate with the algorithm in order to communicate what they don't want to see in their feeds. Nonetheless, P5 expressed frustration that ``they're not learning from my algorithm requests.'' A majority of participants surmise that these individualized solutions, like muting or hiding an ad, yield little success and aren't effective at eradicating weight-loss ads from their feed. P3 explains that even if they hide an ad they find frustrating, they might not see that specific ad again, but they still are delivered similar weight-loss ads.
\begin{displayquote}
I actually did do that hide [option], the one where you can go into three corners and hide [\ldots] It seemed like it didn't matter. Maybe I wouldn't see that exact same thing again, but I feel like things would still pop up. (P3)
\end{displayquote}

When individuals were working on engaging with Intuitive Eating and the broader recovery process, they tried to avoid clicking on or interacting with ads, as they thought that interaction with the ads would trigger delivery of more weight-loss content. 
\begin{displayquote}
I think even just seeing [a weight-loss ad] would make me go down that path a little bit. And then if it was like really bad, then I would click on it and it would be worse. So I learned through that, just don't click on them. Don't interact. (P17)
\end{displayquote}

In addition to muting and blocking, participants sometimes used reporting functions. Some participants saw reporting as an escalation, with greater chances of success at eliminating weight-loss ads from their online experience. Participants also reported ads that felt especially untruthful or unscientific. For instance, P20 reported all weight-loss ads that the algorithm delivered, and indicated that the ads were offensive as the reason for reporting. She theorizes that after repeated reports, the targeting algorithms understood that she didn't want to see weight-loss ads, and subsequently minimized the frequency of such ads. P4 saw reporting as an accessible way to respond to weight-loss ads that could not be verified through scholarly, scientific research. 
\begin{displayquote}
I have muted sometimes but mostly if something is really unhealthy, I try to report it. I think reporting would be the most successful, even though sometimes Instagram and other social media don't really take any actions. Sometimes they do, so I think that's the most successful. (P4)
\end{displayquote}
However, participants expressed that the options for reporting are often unsatisfying and inappropriate for the harm that does occur.  When reporting an ad, users select from a pre-determined list of reasons for reporting. Participants found the reasons provided for reporting limiting and over-simplifying. P11 notes that what is considered offensive is relative based on a given individual’s history with disordered eating. She felt that the reasons for reporting did not match her actual experience -- she felt uncomfortable by weight-loss ads but didn't feel like any of the options accurately described why the ad was distressing. 
\begin{displayquote}
I think in the reporting, being able to just say I don't want to see this instead of it needing to be an objective thing that they're doing wrong. Sure, it might not be offensive for everyone, but it makes me uncomfortable and that should be reason enough. (P11)
\end{displayquote}

Our participants told us that individual strategies have varying levels of success and are neither long-term nor sustainable; for instance, blocking an ad does not prevent future harm and brings little emotional resolve or satisfaction. Some of our participants perceived that some of their actions, such as reporting, are only successful when other people join in.

\subsubsection{Deliberation and paths to collective resistance.}

In addition to the individual strategies described in the section above, our participants engaged in some forms of collective resistance. One strategy was discussing the ads with others including friends and the Intuitive Eating community. Deliberation is central to developing analysis and formulations of the problem \cite{disalvo2009design}. In addition to discussing the harm with their communities, our participants sometimes attempted to reach others who were targeted by the ads by commenting directly on the ad.

\begin{displayquote}
I'll comment [a specific fact] and just hope that somebody sees it, even if it's not the company. If it's just somebody who goes in there and they see that statistic and even if it makes one person be like, ``Huh, I wonder.'' Usually [the ad will] say, ``Scientifically proven to work,'' and I'll be like, ``Can you post the scientific studies?'' where it's something clearly that they claim[\ldots]Some statistic or they'll say, ``Nothing has worked for you, but this will.'' I was like, ``What evidence do you have that this will work for other people?'' Or sometimes I used to comment just, ``Please. If you're interested, look up Health at Every Size and learn the statistics on the failure rate of diets.'' I used to post that a lot and I do do that sometimes on Reddit. (P5)
\end{displayquote}

Overall, participants deployed creative strategies in resisting the harmful effects of weight-loss ads, but were aware of the limitations.

%% file: 5-discussion.tex
\section{Discussion}
Our participants utilized both individual and collective approaches to mitigate the harm perpetuated by targeted weight-loss ads. Individual approaches included blocking ads, reporting, and using ad-blocker browser extensions. Collective approaches included sharing experiences with friends and with the Intuitive Eating community and commenting on the ads to reach other targets. While these approaches offered temporary relief, they are often limited. In order to pursue accountability and scalable change, designers, researchers, and platforms must consider a range of transformative approaches, including what it might look like to refuse targeted advertisements altogether \cite{cifor2019feminist}.

Our results show a cumulative, spatially distributed, mundane and normalized dissemination of harm. Expanding our definition of violence beyond acute and spectacular events helps us recognize the ways in which harm permeates through a multiplicity of mediums, including targeted ads. Slow violence, a concept coined by literary scholar Rob Nixon, is defined as the gradual and cumulative occurrence of violence over time. \cite{nixon2011slow}. Slow violence is violence that is often unseen, delayed and distributed across space and time. Slow violence helps explain how the persistence of sensitive personal data, relevancy models, and engagement dark patterns, significantly impact people with histories of unhealthy body stigma, dieting, and disordered eating over time, even as each individual ad may not seem significantly harmful. Interest in studying the harms of socio-technical systems has grown in CSCW literature in recent years \cite{fox2016exploring,fox2017social, star1999layers, chancellor2019relationships, fox2017imagining, kumar2018uber}. We build on this work and introduce slow violence as a useful framework for the CSCW community to describe the cumulative harms of complex socio-technical systems, such as targeted advertising.

\subsection{The Slow Violence of Harmful Targeted Advertising}
We found that targeted ads deliver harmful content that mirror normative, societal values in the same way that slow violence reifies existing structural and cultural practices of society at large \cite{nixon2011slow}. Embedded within these practices are histories of harm directed particularly towards people with non-normative identities and historically marginalized groups.  Our results showed that weight-loss ads slowly manipulated users to internalize aspirational thinness. Our participants described being unable to escape the idea of ``the perfect body.'' Therefore, the targeted ads created a personalized attack on participants' self esteem. This violence is unique in that it repetitively situates dominant social norms within minute events.

The emphasis on time is the defining factor of slow violence. In conjunction with structural violence, time exacerbates the gradual accumulation of violence, which distinguishes slow violence from other typologies of harm. Slow violence builds off and extends Johan Galtung’s concept of structural violence \cite{galtung1969violence}. Structural violence is violence that exists because of hidden agency and differential power. Nixon’s explicit emphasis on temporality reveals the representational challenges that time presents in delaying destruction and therefore, delaying recognition of the original perpetrators of harm and its victims. Ultimately, slow violence positions time as a masking agent for harm and invites us to use it as a tool to trace previously invisible iterations of harm. Users experience seemingly small, but repeated online interactions -- time is a central component in understanding the cumulative effects of targeted advertising. 

This emphasis on repetition and the passage of time was present in our interviews. 
For instance, P10 struggled with how targeted weight-loss ads acted as a site of replication for societal beauty ideals. P10 described how the persistence of seeing harmful content eroded her ability to resist adopting the thin ideal over time: ``And it just became a snowball effect, where it just builds and builds things until I reached a point where I was convinced that I had to do this.'' (P10) 

\subsubsection{Embeddedness and mundanity make slow violence hard to identify.}
Another major tenant of Nixon's conceptualization of slow violence is mundanity. Type-casted as ``non-events,'' Nixon identifies slow violence events as discrete, as opposed to spectacularized and often widely publicized violent events that draw attention. Harmful targeted ads are seen as non-events within a culture that has normalized targeted ads as part of the typical user's digital experience. Additionally, social media platforms actively work toward  nativity and embeddedness in online advertising; design decisions that some have called dark patterns \cite{an2019recognizing, brignull2003enticing, gray2018dark, mathur2019dark}. This kind of embeddedness hides targeted ads as a site of actual harm since they are typically embedded within other content. Nestled between posts, videos, and articles, online targeted ads almost blend into the background of virtual platforms. A typical targeted ad can look just like any other content that users might come across on their social media feeds. 
Scholar Rik Peeters warns of the effects of embeddedness within algorithmic decision-making. He stresses that embeddedness limits transparency and pathways for establishing accountability \cite{peeters2020agency}. 

Participants in our interviews initially experienced difficulty recalling their experiences with targeted ads since the ads felt so ubiquitous and nearly unnoticeable. The prompts that we used were helpful in reminding them of their experiences with the ads. Since platforms display targeted ads among other content, participants claimed that it feels less noticeable and harder to define. Our participants described how the embeddedness produces a naturalizing effect, which makes it both difficult to identify and mitigate harm. 

\begin{displayquote}
So the thing about ads, right, is that also I have a tendency to be like, ``Oh, they don't matter, I don't look at them,'' but it's not true. We see thousands of ads a day, but nonetheless, the repetition brands your brain, like wires your brain with, ``Oh, well this image is sexy, so that's what sexy looks like.'' (P1) \end{displayquote}

Slow violence draws our attention to the unremarkable nature of targeted advertising as potentially distracting from the harm that it causes. Ultimately, participants were able to recognize targeted weight-loss ads as harmful but communicating them as such was still a difficult task. 
\subsubsection{Naming slow violence produces visibility and a path to accountability.}
The act of naming slow violence brings visibility to the harms and experiences of marginalized groups, and brings attention to their nuanced needs. Scholar Brydolf-Horwitz has proposed that naming slow violence produces accountability as well as potential resources for marginalized people. She argues that ``creating language around experiences that have been invisible and difficult to address is a critical step in generating forms of accountability, as well as support and resources for those targeted'' \cite{brydolf2018embodied}. 

Our data clearly illustrates a pattern of recognition of violence among our participants. This is evidenced in the individual mitigation strategies they employed such as reporting the ad to the platform and downloading third-party ad blockers. We saw evidence of witness and resistance as many participants used hypothetical situations to recount how they would have protected themselves and other people they recognized as vulnerable, such as women and children. 

Differential visibility within slow violence is further explored by scholar Thom Davies in his ethnographic study of minority responses to pollution in the Louisiana Cancer Valley. He suggests that the subjects of slow violence, such as the victims of pollution, are fully aware of the harm they are enduring and are the best whistle blowers for violence that tends to evade public scrutiny \cite{davies2019slow}. Similarly to Davies, we rely on the experiences and situated knowledge of our participants as our primary approach for capturing and naming violence. 

We highlight the ways in which they resist harm and we use slow violence as a framework for making their experiences and needs visible. Our research speaks directly to a question Nixon poses: ``How do we make slow violence visible yet also challenge the privileging of the visible?'' \cite{nixon2011slow}. The spatial distribution of targeted ads uniquely positions certain users as vulnerable. As targeted ads proliferate both public and private spaces, calls for inclusive visibility become a challenge that we will pursue in future work. 

\subsection{Future Work}

It is important to note that our study only interfaced with participants at a single moment in time. Analyzing longitudinal changes and developments will clarify how algorithmic harm manifests over time. Overall, we hope that our research opens an avenue for future work addressing online harms that may be less apparent, as articulated by the concept of slow violence. We propose two main areas around future work: designing for user agency and public policy interventions.

We hope that this work inspires design solutions for collective approaches to this issue. During interviews, participants proposed creative, collective technical solutions, such as community-supported ad blockers, more granular ad settings, and crowdsourced content warnings. This research opens an avenue for future work to design opportunities for users to exercise agency. While weight-loss ads are one example of sensitive advertising, this line of research can expand to analyze the harms of ads that target other sensitive health data, such as personal data indicating fertility status, alcoholism, addiction, and more. Effective approaches to this issue will only be mobilized from organizing the people most vulnerable to advertising of sensitive topics.

This line of research aims to inspire public policy interventions, specifically around regulating targeting practices and limiting data gathering. In 2022, lawmakers in the United States introduced the Banning Surveillance Advertising Act, which prohibits online advertisers from targeting ads to individuals based on their personal information, including data that identifies a person as part of a protected class \cite{congress_2022}. While we are supportive of public policy curbing targeting practices, we also hope that our research complicates how targeted advertising can harm vulnerable people beyond the traditional category of a protected class. Beyond public policy, we've already seen platforms begin to self-regulate their own targeting practices. In July 2021, Pinterest, a popular social media platform that our participants reported hosted harmful ads, updated their policies to prohibit all weight-loss ads \cite{pinterest_newsroom_2021}. We hope that platforms and advertisers respond to the expressed needs of communities calling for the removal of weight-loss ads. Not only is it important to restrict targeting practices, but we also hope our research encourages public policy that regulates what kind of sensitive data can be gathered. While the General Data Protection Regulation (GDPR), the European Union's legal framework for data protection, has defined regulations around ``data concerning health,'' most of the protections around this sensitive data concerns data processing in healthcare settings. Our research hopes to inspire broader protections for sensitive data gathering and processing in the targeted advertising space.

%% file: 6-conclusion.tex
\section{Conclusion}

In this paper, we study how targeted weight-loss ads harm individuals with histories of unhealthy body stigma, dieting, and disordered eating. Our results show that participants identified targeted delivery as harmful in three major ways: the persistence of personal data that can expose vulnerabilities; over-simplifying algorithmic relevancy models; and design patterns encouraging engagement that can facilitate unhealthy behavior. Participants worked to make sense of the harms of targeted weight-loss ads, such as by recognizing their pseudoscientific claims. While participants changed their online behaviors to diminish the harms of targeted weight-loss ads, they also recognized the limitations of available mitigation strategies. We use the concept of slow violence to analyze how the gradual and seemingly innocuous accumulation of exposure to weight-loss content can cause jarring emotional and physical harms. Finally, we propose working with harmed communities to develop design and policy interventions that mitigate the harms of targeted ads, particularly to those most vulnerable to them.